# A complete photoluminescence polarization palette from a photonic-crystal slab

**A.V. Peretokin[1,*], M.V. Stepikhova[1], N.A. Gippius[2], A.A. Bogdanov[3,4], S.G. Tikhodeev[5,6], R.K. Nazarov[4], M.V. Shaleev[1], A.V. Novikov[1] and S.A. Dyakov[2]**

[1]Institute for Physics of Microstructures RAS GSP-105, Nizhny Novgorod 603950, Russia

[2]Skolkovo Institute of Science and Technology, Nobel Street 3, Moscow 143026, Russia

[3]Harbin Engineering University, Qingdao Innovation and Development Center, Sansha Road 1777, Qingdao 266404, Shandong, China

[4]ITMO University, Birjevaja Line V.O. 14, St. Petersburg 199034, Russia

[5]Lomonosov Moscow State University, Leninskie Gory, GSP-1, Moscow, 119991, Russia

[6]Prokhorov General Physics Institute of the Russian Academy of Sciences, Vavilova St. 38, Moscow, 119991, Russia

*aperetokin@ipmras.ru;

**Abstract**:

Integrated light sources with tailored polarization are essential for integrated photonics, optical communication, sensing, and structured-light applications. However, generating a broad set of polarization states from spontaneous emission usually relies on structures with broken mirror, rotational, or inversion symmetries, or on emitters with predefined polarization properties. Here, we demonstrate a complete photoluminescence polarization palette in a single achiral photonic-crystal slab with a hexagonal lattice of holes and embedded self-assembled Ge(Si) nanoislands. We show that the polarization of the emitted light is governed by the interplay among modes with different symmetries rather than by the properties of unpolarized emitters. Along the high-symmetry directions of the Brillouin zone, symmetry enforces purely linear emission and enables direct polarization-based classification of the photonic-crystal modes. Away from these directions, all Stokes parameters become non-zero, providing full-Stokes control of photoluminescence and highly polarized emission. We further demonstrate polarization vortices associated with symmetry-protected and Friedrich-Wintgen bound states in the continuum, as well as radiating polarization singularities not associated with bound states in the continuum. Through appropriate geometric optimization, the same platform supports photoluminescence with a full polarization palette as well as polarization vortices. Our results establish achiral photonic-crystal slabs as monolithic sources of symmetry-programmable photoluminescence polarization.

## Introduction

Polarization is a fundamental property of electromagnetic waves and a key degree of freedom in modern optics and photonics.[1] Precise polarization control has enabled major advances in imaging, spectroscopy, optical communications, quantum optics, sensing, and light–matter interaction. For decades, this control has been successfully achieved using conventional optical components, including wave plates, polarizers, dichroic filters, and spatial light modulators.[2–5] These approaches provide robust and accurate polarization manipulation and remain indispensable in free-space optical systems. However, their operation often relies on propagation through bulk optical elements or externally configured modulation schemes, which makes further miniaturization and dense integration with on-chip light sources challenging. At the same time, light is inherently vectorial.[6] Its polarization state is conveniently described by the Stokes parameters, which quantify the relative contributions of linear, diagonal, and circular polarization components and therefore provide a complete experimental description of the polarization state. In structured optical fields, the Stokes parameters can vary in real or momentum space, giving rise to polarization singularities.[7–10] The most common singular structures are lines of linear polarization, known as L-lines; points of circular polarization, known as C-points; and vector singularities, or V-points, around which polarization vortices are formed.

All-dielectric metasurfaces and photonic-crystal slabs have emerged as powerful platforms for the development of sources of circularly polarized radiation,[11–15] including lasers ,[16–19] as well as for realizing and controlling polarization singularities in compact photonic structures.[8,20–38] Of particular importance are bound states in the continuum (BICs): nonradiative eigenmodes embedded in the radiation continuum whose coupling to outgoing waves is suppressed by symmetry or destructive interference, resulting in, theoretically infinite radiative Q factor. In momentum space, BICs appear as topological polarization singularities, or V-points, around which the far-field polarization forms vortices with quantized topological charges.[39–46] C-points can also be engineered in the vicinity of BICs, for example through symmetry reduction that transforms or splits the original vector singularity,[47–51] as well as in more complex architectures, including chiral and bilayer photonic structures.[22,27,36,52–55] These developments have established polarization-singularity control and topological-charge engineering as central directions in modern nanophotonics.

Several studies have addressed polarization engineering in highly symmetric photonic structures. In platforms with preserved structural symmetry, polarization-selective modal responses have been demonstrated, although the connection between the observed polarization properties and the underlying modal symmetries has not always been analyzed explicitly.[35,56–58] However, generating a broad set of polarization states from spontaneous emission usually relies on structures with broken mirror, rotational, or inversion symmetries, or on emitters with predefined polarization properties.[59] Alternative routes, including twisted bilayer metasurfaces[54,60,61] and magneto-optical photonic structures,[38] can provide access to a wide range of polarization states, including coverage of the Poincaré sphere, but at the cost of increased structural complexity or additional material functionality. Highly symmetric structures capable of supporting multiple polarization states have also been reported; however, these demonstrations have relied either on transmission schemes with complex incident fields,[35] or on magnetic responses.[38] Thus, a key unresolved challenge is to realize a single achiral photonic platform that can generate the full range of polarization states from spontaneous emission, spanning purely linear, circular, or more complex polarizations, through the intrinsic symmetry and modal composition of the photonic structure itself.

In this work, we demonstrate a single achiral photonic-crystal slab with a hexagonal lattice of air holes and $\boldsymbol{C}_{6v}$ symmetry as a platform for generating a complete photoluminescence polarization palette from self-assembled Ge(Si) nanoislands acting as unpolarized spontaneous-emission sources. We show that the emitted polarization is governed by the interplay among photonic-crystal modes with different symmetries, rather than by the intrinsic polarization properties of the emitters. This enables purely linear polarization along the high-symmetry directions of the Brillouin zone, circularly polarized states away from these directions, and polarization vortices associated with both BICs and radiating singularities at which the polarization becomes undefined. We further show that a BIC, acting as a V-point, can simultaneously serve as the center of two polarization vortices with the same topological charge but orthogonal polarization states. Such symmetry-programmable control of photoluminescence in a single achiral structure provides a route toward monolithic on-chip sources with engineered polarization states, relevant for classical and quantum communications,[62–64] optical manipulation,[65] sensing,[66] vortex lasing,[31] biomedicine,[67] and other applications of advanced photonics.[68–70]

## Results and discussion

### Design and fabrication

In this work, we experimentally investigate a photonic crystal slab (PCS) with a hexagonal lattice of air holes. The PCS has the following geometric parameters: lattice period $a$=575 nm, holes' diameter $d$=277 nm, etching depth $h_{etch}$=250 nm, Si waveguide layer thickness $h_{Si}$=335 nm, and $SiO_2$ layer thickness $h_{SiO2}$=2 μm. The conditions for the growth of the initial SiGe structure and the formation of the sample are described in more detail in the Samples and Methods section. Figure 1a presents a schematic representation of the investigated sample. Figure 1b shows the emission spectrum of a structure with Ge(Si) nanoislands outside the PCS (gray curve). As can be seen, the nanoislands emit at room temperature in the near-infrared wavelength range 1.2-1.6 μm (800-1000 meV), which corresponds to the telecommunication O- and C-bands. This PCS was selected based on the results of our previous studies on the effect of PCS geometry on its band structure and luminescence response.[71,72] With these PCS parameters, several modes with symmetry-protected bound states in the continuum (SP-BICs) at the Γ-point, a mode with flat dispersion near the Γ-point, and a radiating mode are observed in the radiation range of Ge(Si) nanoislands simultaneously.

### Polarization states of PCS radiation: from linear to circular

Figure 1b presents the photoluminescence (PL) spectra of the investigated PCS and the initial (as grown) sample, both measured using a standard micro-PL setup. As can be seen from the obtained spectra, the PCS spectrum shows a nearly two-orders-of-magnitude increase in the peak PL intensity (~94 times) and a 6.5-fold enhancement of the integrated PL intensity compared with the initial sample. This PL enhancement is attributed to the coupling between Ge(Si) nanoislands' radiation and the PCS modes.

In the spectrum, we identify seven peaks corresponding to the PCS modes at the Γ-point of the Brillouin zone. According to group theory, the investigated PCS belongs to the symmetry point group $\boldsymbol{C}_{6v}$. A family of modes of the PCSs with such symmetry have 8 eigenmodes at the $\Gamma$-point: $A_1$, $A_2$, $B_1$, $B_2$, $E_1^{low}$, $E_1^{up}$, $E_2^{low}$, and $E_2^{up}$.[73,74] In our PCS, all of these eigenmodes, except $A_1$, can be observed in the energy range 800-1000 meV. Mode $A_1$ is invisible because its energy is outside the emission range of the Ge(Si) nanoislands.

Figure 1c shows theoretically calculated emissivity (Stokes parameter $S_0$) band structure near the $\Gamma$-point for the studied PCS. The corresponding modes at the $\Gamma$-point are labeled in Figure 1b. Away from the $\Gamma$-point, along the high-symmetry directions $\Gamma \to M$ and $\Gamma \to K$, the modes are described using irreducible representations of $\boldsymbol{C}_{1v}$ point group. Along these directions, group theory predicts two distinct types of modes, $A$ and $B$, which differ in their parity.[75,76]

At the same time, the polarization of light emitted in the high-symmetry directions $\Gamma \to M$ and $\Gamma \to K$ is strictly defined. Along these directions, the polarization can be only horizontal ($p$-polarized) or vertical ($s$-polarized). This restriction arises because the high structural symmetry requires that the polarization must be preserved under a 180° rotation, a condition not satisfied for any other polarization states. Moreover, the mode symmetry and the polarization of the emitted light are linked with each other. Indeed, the modes with $A$-symmetry can emit only $p$-polarized light (Stokes parameter $S_1>0$) light, whereas the $B$-symmetrical modes are $s$-polarized in the far field (Stokes parameter $S_1<0$) (see Figures 1e and 1f). Based on the mode's symmetry in the $\Gamma \to M$ and $\Gamma \to K$ directions, and their polarization states in the far-field, the modes can be divided into 2 groups: modes with the same parity in the $\Gamma \to M$ and $\Gamma \to K$ directions ($A_1$, $A_2$ and radiation modes $E_1$) and modes with different parity in these directions ($B_1$, $B_2$ and $E_2$) (see Figure 1c-e). This classification is crucial for understanding the relationship between the topological charges of BICs and the symmetry of the mode on which they arise. We will discuss this in more detail in the section devoted to the polarization vortices generated by the PCS.

As a result, in the high-symmetry directions $\Gamma \to M$ and $\Gamma \to K$, only Stokes parameter $S_1$ is nonzero (it can be shown that the Stokes parameters $S_2$ and $S_3$ in these directions are negligible – see Figure SM1 in the Supplementary materials). Such polarization characteristics are known as L-lines.[36] Furthermore, it is important to note that in resonances, the degree of polarization (DoP), defined as:

$$DoP = \sqrt{\frac{S_1^2 + S_2^2 + S_3^2}{S_0^2}} \qquad (1)$$

is close to unity (Figure 1g).

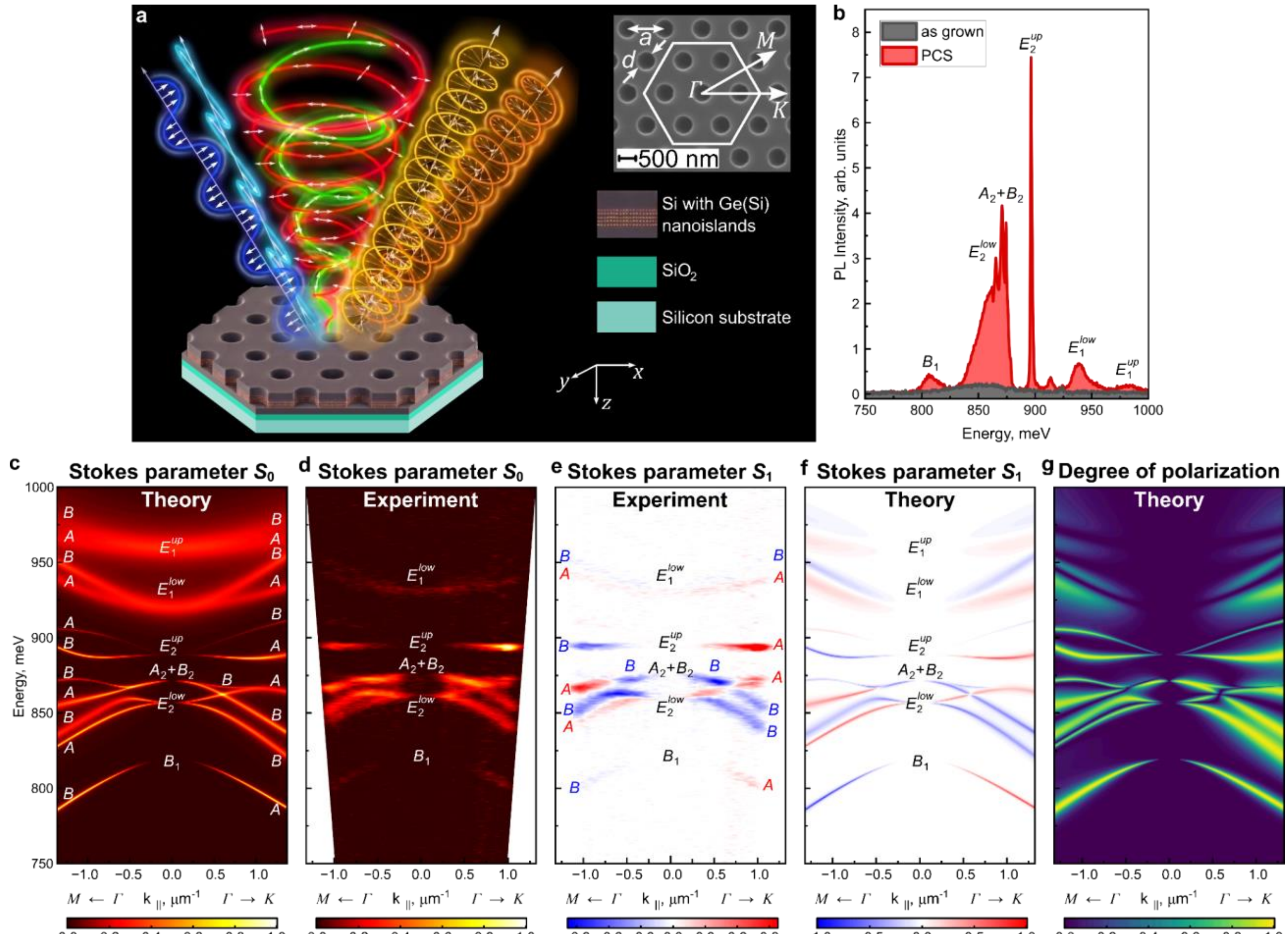

Figure 1 – (a) Top: Schematic representation of the photonic crystal slab under study, characterized by a complete palette of radiation polarization. Linear polarization of PCS radiation is shown in blue and cyan, polarization vortices in red and green, and circular polarization in yellow and orange. Top right: SEM-image of the studied PCS, where $a$ is the lattice period, and $d$ is the diameter of the air holes. The first Brillouin zone of PCS is shown by white line. The white arrows in the figure mark the high-symmetry directions $\Gamma \rightarrow M$ and $\Gamma \rightarrow K$.

(b) PL spectra of the PCS (red curve) and the initial (as grown) sample on which the PCS was formed (gray curve). The spectra were measured with a collection angle $2\alpha = 30°$.

(c) The band diagram of PCS emissivity near the Γ-point of the first Brillouin zone, theoretically calculated for the Stokes parameter $S_0$.

(d) and (e) Band diagrams of the PCS emissivity, experimentally measured for the Stokes parameter $S_0$ (d) and the Stokes parameter $S_1$ (e). The collection angle in measurements was chosen to be $2\alpha \approx 1.5°$, the angular step $\Delta\vartheta \approx 0.75°$. More detailed information on these measurements and the data of the band diagram measurements carried out for PCS emissivity in the *s*- and *p*-polarizations are given in Figure SM2 in the Supplementary materials.

(f) and (g) Numerically simulated PCS's emissivity band structure for the Stokes parameters $S_1$ (f), and the degree of polarization (g) along the high-symmetry directions $\Gamma \rightarrow M$ and $\Gamma \rightarrow K$ near the $\Gamma$-point. In panels (b)-(f), the letters denote the PCS modes. The labels $B_1$, $E_2^{low}$, $A_2$, $B_2$, $E_2^{up}$, $E_1^{low}$, and $E_1^{up}$ correspond to the modes of $\boldsymbol{C}_{6v}$ symmetric PCS at the Γ-point. In panels (c) and (e) the labels $A$ and $B$ for the modes far from the $\Gamma$-point correspond to irreducible representation of the $\boldsymbol{C}_{1v}$ symmetry point group.

As one moves away from the $\Gamma$-point and the high-symmetry directions $\Gamma \rightarrow M$ and $\Gamma \rightarrow K$, the symmetry of the structure decreases. Consequently, radiation with non-zero Stokes parameters $S_2$ and $S_3$ may occur. Figure 2a shows the calculated maps of the Stokes parameters, the degree of polarization, and the polarization ellipses. The simulations were performed in the space of wavevectors $k_x$, $k_y$, over the range from −1.5 to 1.5 μm$^{-1}$ in both coordinates. To construct these maps, the eigenfrequencies of the lower branch of the $E_2^{up}$ mode were determined for each point in $k$-space. Then, for a given point in $k$-space, all the Stokes parameters were calculated for the determined eigenfrequency. The maps of the Stokes parameters $S_1$-$S_3$ were then normalized to the Stokes parameter $S_0$.

For the map of polarization ellipses, we calculated the lengths of the long ($a$) and short ($b$) semi-axes:

$$a = \sqrt{\frac{I_p + \sqrt{S_1^2 + S_2^2}}{2}}, b = \sqrt{\frac{I_p - \sqrt{S_1^2 + S_2^2}}{2}}, \tag{2}$$

where $I_p$ is the polarized part of emission:

$$I_p = \sqrt{S_1^2 + S_2^2 + S_3^2},$$

and the ellipse orientation angle $\psi$ (ellipse orientation) is given by:

$$\psi = \frac{1}{2} \arg\left(S_1 + i \cdot S_2\right). \tag{3}$$

In full agreement with Figure 1, the data in Figure 2a show that the Stokes parameters $S_2$ and $S_3$ vanish along the high-symmetry directions $\Gamma \rightarrow M$ and $\Gamma \rightarrow K$, but become non-zero elsewhere. Away from these directions, the radiation is no longer purely $s$- or $p$-polarized and acquires a partial circular polarization component. As seen in Figure 2a, the degree of polarization is close to unity in the vicinity of each mode, and is close to zero at all other points in $k$-space and at the $\Gamma$-point. The zero degree of polarization corresponds to an undefined polarization state of the emitted light, which can arise either from the absence of a distinguished symmetry direction in the $\boldsymbol{C}_{6v}$-symmetric photonic crystal (at the $\Gamma$-point) or from the unpolarized nature of emitters (at all other points).

To validate these theoretical predictions, we measured the Stokes parameters of the PCS luminescence response in the selected direction, outside the $\Gamma$-point and away from the high-symmetry directions $\Gamma \rightarrow M$ and $\Gamma \rightarrow K$. The measurements were performed at detection angles
$\vartheta \approx 11.5°$ and $\varphi = -20°$ ($\vartheta$ is the polar angle; $\varphi$ is the azimuthal angle, see Figure SM3 in Supplementary materials for more details), with a collection angle $2\alpha \approx 3°$. These results are shown in Figure 2b.

The experimental data confirm that outside the high-symmetry directions, the PCS luminescent response is indeed characterized by non-zero Stokes parameters, in good agreement with the numerical simulations shown in Figure 2a. In the upper panel of Figure 2b, blue, green and red areas represent the partial contributions to the total emission $S_0$ from the polarized components with the Stokes parameters $S_1$, $S_2$, and $S_3$, respectively. Three lower panels in Figure 2b, show the corresponding experimentally measured Stokes parameters $S_1$, $S_2$, and $S_3$. The spectra of the measured polarization components (vertical, horizontal, diagonal, right- and left-circular) are provided in Figure SM4, of the Supplementary materials. The boundary between the gray and red regions indicates the overall degree of polarization. The maximum experimentally achieved degree of polarization DoP=0.94 obtained for the lower branch of $E_2^{up}$ mode (energy ~895 meV). As in theoretical predictions, the degree of polarization reaches its maximum value at resonances. It is worth noting once again that the radiation of Ge(Si) nanoislands is not intrinsically polarized; all observed polarization effects arise directly from the PCS modes.

We also experimentally measured the map of the Stokes parameter $S_3$ for the lower branch of the $E_2^{up}$ mode (energy ~895 meV). The results are shown in Figure 2c (upper panel) alongside with the corresponding theoretical simulations (upper panel). Qualitatively good agreement is observed between experiment and theory. Some discrepancies can be attributed to the relatively large collection angle of the PL signal ($2\alpha \approx 3°$), and to inaccuracies in the diaphragm positioning, which affect the determination of the detection angle. In the lower panel of Figure 2c, we show the size of the aperture in $k$-space with a solid violet circle, the error in determining the position of the aperture is shown by a dashed ellipse. The reduced error of the azimuthal angle determination is $\delta\varphi = 4\%$, and that of the polar angle is $\delta\vartheta = 30\%$. In the figure, the error in the polar angle is shown along the $k_x$ axis, and that of the azimuthal angle along the $k_y$ axis. For more information, see Supplementary materials. Theoretically, we observe a lower degree of circular polarization than in the experiment, which may be explained by slight differences between the parameters of the modeled structure and those of the actual PCS under study.

Importantly, the polarization of the observed radiation can be controlled by changing the detection angle. Figure 2d shows the normalized Stokes parameters measured at different azimuthal angles: $\varphi = -20°$ (orange curves) and $\varphi = +15°$ (violet curves), with the polar angle $\vartheta \approx 11.5°$ in both cases. Experimental results clearly show changes in both the sign and magnitude of the Stokes parameters, particularly for the lower branch of the $E_2^{up}$ mode (energy ~895 meV). The experimentally obtained maximum values of the normalized Stokes parameters are $|S_1/S_0| = 0.87$, $|S_2/S_0| = 0.96$, and $|S_3/S_0| = 0.54$.

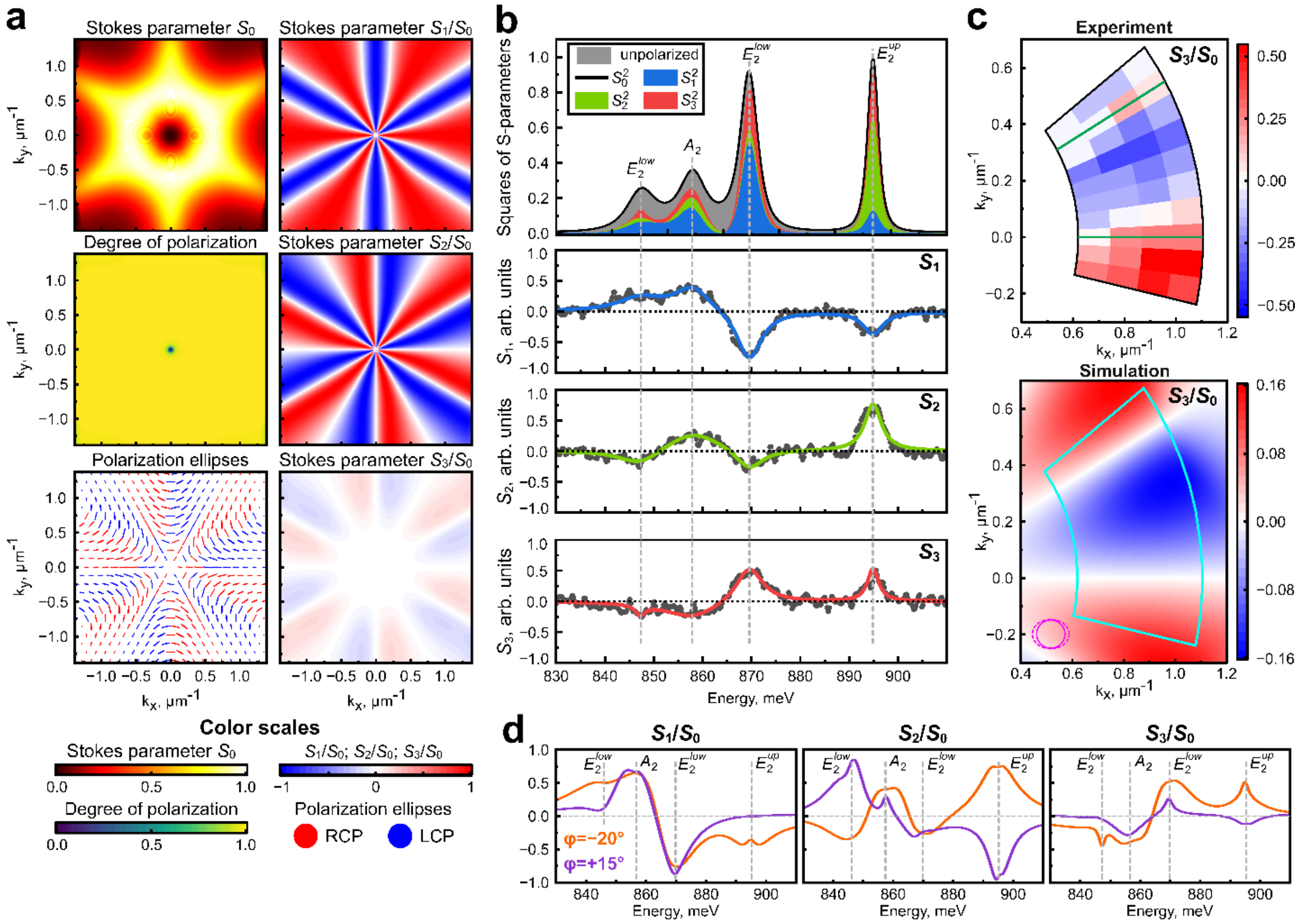

Figure 2 (a) Numerically simulated maps of the normalized Stokes parameters $S_0$, $S_1/S_0$, $S_2/S_0$ and $S_3/S_0$, the degree of polarization and polarization ellipses obtained for the emissivity of the lower branch of the $E_2^{up}$ mode. In the map of polarization, the red color of the ellipses indicates right circular polarization (RCP), and the blue color indicates left circular polarization (LCP). The ellipse sizes are proportional to the intensity of the polarized part emissivity. The color scales for panel (a) are shown below them.
(b) Experimentally measured Stokes parameters of the PCS luminescence response in the selected direction, outside the $\Gamma$-point and away from the high-symmetry $\Gamma\to M$ and $\Gamma\to K$ directions. The measurements were performed at azimuthal angle $\varphi = -20°$ and polar angle $\vartheta \approx 11.5°$, the collection angle was $2\alpha \approx 3°$. Three lower panels show the experimentally measured Stokes parameters $S_1$, $S_2$, and $S_3$; in all panels, gray dots are experimental data and solid curves are Lorentzian fit. The upper panel shows the squares of the experimentally observed Stokes parameters (Lorentzian fits), where the blue, green and red areas correspond to the partial contribution to the total emission ($S_0$) from the polarized components with the Stokes parameters $S_1$, $S_2$, and $S_3$, respectively. The boundary between the gray and red regions indicates the overall degree of polarization. In all panels, vertical gray dotted lines mark the PCS modes; their labels correspond to the group-theory mode notations at the Γ-point. Additional details on the Stokes parameters measurements are provided in Figures SM3 and SM4 of the Supplementary materials.
(c) Maps of experimentally measured (upper panel) and theoretically calculated (lower panel) normalized Stokes parameter $S_3/S_0$. The green lines in experimentally map indicate the high-symmetry directions $\Gamma\to K$ (horizontal line) and $\Gamma\to M$ (inclined line). The cyan contour in the lower panel shows the region in $k$-space where the measurements were carried out. Violet circles correspond to the aperture used in the experiment; the solid circle shows the nominal size of the aperture, and the dashed circle indicates the size of the aperture including the measurement error.
(d) Normalized Stokes parameters experimentally measured at different azimuthal angles: $\varphi = -20°$ (orange curves) and $\varphi = +15°$ (violet curves) with the same polar angle $\vartheta \approx 11.5°$ in both cases. In all figures, vertical gray dotted lines indicate modes corresponding to the peaks observed in the PL spectra. Experimentally measured Stokes parameters for the azimuthal angle $\varphi = +15°$ are shown in Figure SM5 in Supplementary materials.

In the experimentally studied structure, the degree of circular polarization is 0.53. By choosing an appropriate geometry of $\boldsymbol{C}_{6v}$-symmetric PCS , it is possible to obtain a radiation pattern where one of emission directions is characterized by the degree of circular polarization close to unity; such a direction is referred to as a C-point. Using a genetic algorithm,[77,78] we conducted optimization of the PCS geometric parameters and achieved the degree of circular polarization exceeding 0.97. The PCS parameters used in simulations are: $a$=856 nm; $d/a$=0.45; $h_{Si}$=284 nm; $h_{etch}/h_{Si}$=0.73; $h_{oxide}$=2910 nm. With these PCS parameters at an energy of 900 meV, we observe six pairs of C-points arranged symmetrically relative to the $\Gamma\to M$ directions, which, as noted above, are L-lines. These C-points are localized in $k$-space at wave vectors $k_C = \left(k_x = 1.338\ \mu\text{m}^{-1}; k_y = \pm 0.553\ \mu\text{m}^{-1}\right)$, as well as symmetry-equivalent points, as can be seen from the map of polarization ellipses (Figure 3g). At these points, the linear polarization vanishes ($S_1$=$S_2$=0), and the total degree of polarization reaches unity, as shown in Figures 3b-d and 3f.

As is known, C-points are characterized by the presence of polarization vortices around them, with half-integer topological charge[36], which is determined as:

$$q = \frac{1}{2\pi} \oint_C dk \cdot \nabla_k \psi(k), \qquad q \in Z \tag{4}$$

where $\psi$ is the "ellipse orientation", defined in formula (3). The polarization vortex around the C-point is illustrated in Figure 3h for $k_C = (k_x = 1.338\ \mu m^{-1}; k_y = 0.553\ \mu m^{-1})$, in which we observe right circular polarization emission (see Figures 3d and 3g). It can be seen from Figure 3h, that the main axis of the polarization ellipse rotates half a turn clockwise, while going around the C-point counterclockwise. This means that we are observing a C-point with a topological charge of $q = -1/2$.

In the genetic-algorithm optimization procedure, we used the ratio $S_3/S_0$ as a figure of merit (FoM). We also limited the geometry of the PCS (period, holes diameter and etching depth, thickness of layers) based on practically feasible parameters. We searched for C-points at an energy of 900 meV and limited to azimuthal angles $\varphi = 0 \div 30°$ (due to the symmetry of the system) and polar angles of $\vartheta = 0 \div 20°$ (based on the possibilities of further experimental observations). For the ideal C-point, the corresponding FoM value is $|S_3/S_0| = 1$, while in our case we achieved $|S_3/S_0| = 0.967$. Thus, the states with a high degree of circular polarization that we theoretically observe in Figure 3 are C-points: we observe them in pairs (LCP and RCP), they are located mirror-symmetrically relative to L-lines ($\Gamma \rightarrow M$ directions), and we also observe a vortex with a half-integer topological charge around them.

To demonstrate that this set of parameters is not unique for observing C-points in $\boldsymbol{C}_{6v}$ symmetric structures, using a genetic algorithm, we found another set of PCS parameters that also allows us to observe six pairs of C-points. The results of these calculations along with the corresponding PCS parameters are presented in Figure SM6 in the Supplementary materials.

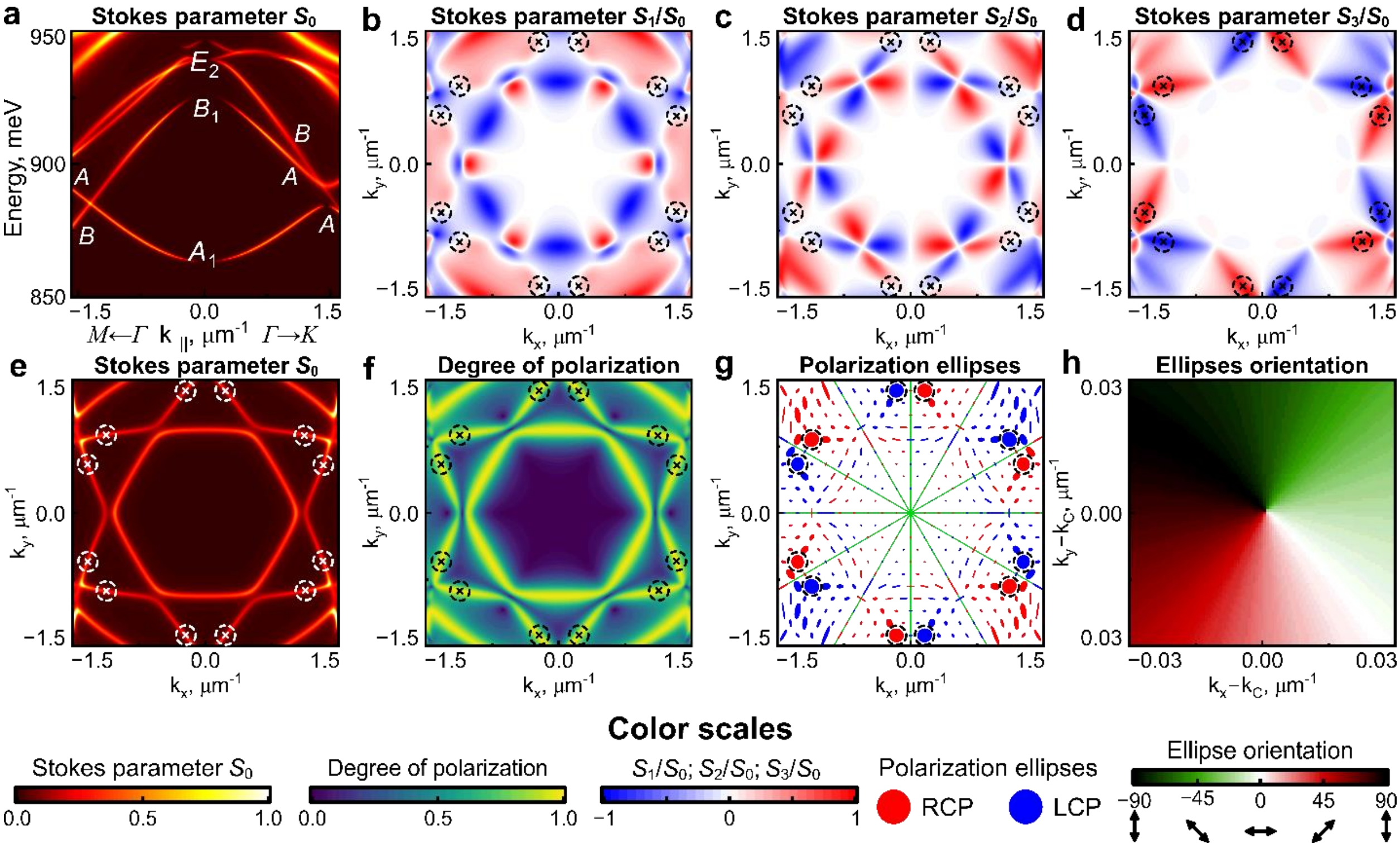

Figure 3 (a) Theoretically calculated emissivity band structure for the Stokes parameter S0 obtained for PCS with the optimized parameters; (b)-(e) Maps of (b) the normalized Stokes parameters; (f) the degree of polarization; (g) the polarization ellipses (h) the orientation of polarization ellipses near the C-point. Panels (b)-(h) are calculated at an energy of 900 meV. The ellipse sizes are proportional to the intensity of the polarized emission. The dashed circles with crosses in the panels (b)-(f) and dashed circles in the panel (g) indicate the C-points. Green lines on the panel (g) correspond to the high-symmetry $\Gamma \to M$ and $\Gamma \to K$ directions, which are L-lines. The ellipse orientation shown in panel (h) is constructed for a C-point localized in $k$-space at a wave vector $k_C = (k_x = 1.338\ \mu m^{-1}; k_y = 0.553\ \mu m^{-1})$.

## Generation of polarization vortices in the PCS

The polarization states described above are observed at specific wave vectors. However, when considering the far-field PCS emission as a whole, other interesting polarization features also emerge. In particular, this concerns BICs. As is well known BICs have a topological nature: a polarization vortex is formed in the far field around a BIC, and the BIC itself is characterized by a topological charge $q$.[39]

In this work, we experimentally investigated polarization vortices formed around symmetry-protected BICs (SP-BICs) for PCS modes $E_2^{up}$, $A_2$, and $E_2^{low}$ modes, which are localized at the Γ-point of the Brillouin zone. Figure 4 shows the experimental measurements of the PCS band structure (panel a) and polarization vortices localized around the SP-BICs of $E_2^{up}$ (the most intense lower branch, panel b), and $A_2$ modes (panel c). Polarization spirals in this Figure illustrate the topological charge $q$ of the SP-BICs and the mode polarization observed when scanning around the Γ-point (panels d and e). An additional figure explains the polarization spirals (panel f). The band structure measurements were carried out with a signal collection angle $2\alpha = 1.5°$, and no polarizing elements were used in the setup for this measurement. For the polarization vortex measurement around SP-BICs, a collection angle $2\alpha = 3°$ was used, and we added a polarizer in front of the entrance lens of the spectrometer.

To experimentally measure polarization vortices, we moved away from the Γ-point ($\vartheta = 0°$) by an polar angle $\vartheta = 7.3°$, and fixed it for all time of measurements. Then we changed the azimuthal angle $\varphi$ by steps of 30° and traversed the points located along the high-symmetry directions Γ→K (azimuthal angles $\varphi$ are equal to 0°, 60°, 120°, 180°, 240°, and 300°) and Γ→M (azimuthal angles $\varphi$ are equal to 30°, 90°, 150°, 210°, 270°, and 330°) (see Figure 4f, panel "Azimuthal angle"). This gives us a set of 12 points arranged on a circle around the $\Gamma$-point. At each point, a series of spectra was measured as a function of the polarizer rotation angle. The resulting spectra were analyzed for peaks associated with the PCS modes $E_2^{up}$ (lower branch), $A_2$, and $E_2^{low}$ (upper and lower branches). For each of them, the peak intensity was plotted as a function of the polarizer rotation angle in polar coordinates, which made it possible to construct polarization figure-eights. By constructing such figure-eight polarization plots for the selected peaks at all 12 points, we obtained the images of polarization vortices shown in Figures 4b and 4c (for details, see also Figure SM7 in the Supplementary materials). These figures show the measurement results of the polarization vortices around the SP-BICs inherent in modes $E_2^{up}$ and $A_2$.

To estimate the topological charge, we examine the behavior of polarization vortices around the SP-BICs. For the $E_2^{up}$ mode (Figure 4b), at the point with azimuthal angle $\varphi = 0°$ ($\Gamma \to K$ direction), we observe a horizontal polarization figure-eight, with the an ellipse orientation of $\psi = 0°$. At $\varphi = 30°$ ($\Gamma \rightarrow M$ direction), the ellipse orientation is $\psi = -60°$, while at $\varphi = 60°$ (next $\Gamma \rightarrow K$ direction), the ellipse orientation becomes $\psi = -120°$, and this trend continues for subsequent angles. Accordingly, upon completing a full closed loop around the BIC (i.e., returning to the initial azimuthal angle $\varphi = 0°$ after 360°), the ellipse orientation rotates by $\Delta\psi = -720° = -4\pi$ relative to its starting point ($\varphi = 0°$). The topological charge can thus be experimentally determined as:

$$q = \frac{\Delta\psi}{2\pi}, \tag{5}$$

where $\Delta\psi = \psi|_{\varphi=2\pi} - \psi|_{\varphi=0}$ – the total rotation angle of the polarization upon traversing a closed contour around the BIC. Thus, for the SP-BIC of the $E_2^{\mathrm{up}}$ mode we obtain $q = -4\pi/2\pi = -2$. For the SP-BIC of the $A_2$ mode, the topological charge turns out to be $q = +1$, since $\Delta\psi = 2\pi$, and the direction of rotation of the polarization figure-eight coincides with the direction of the polar angle traversal (Figure 4c).

The rotation of polarization around the BICs can be more clearly observed using polarization spirals. In Figure 4d, the polarization spirals are shown for the SP-BICs of $E_2^{\mathrm{up}}$ and $A_2$ modes. For topological charge $q = -2$ (red spiral, mode $E_2^{\mathrm{up}}$) the spiral winds clockwise and makes two full turns. For topological charge $q = +1$ (violet spiral, mode $A_2$) the spiral winds counterclockwise and makes one full turn. Thus, the number of the spiral turns indicates the magnitude of the topological charge $|q|$, and the direction of the spiral's unwinding indicates its sign: the counterclockwise direction corresponds to positive $q$, and the clockwise direction to negative $q$.

Here we should also pay attention to the behavior of vortices around the SP-BIC of the $E_2^{\mathrm{low}}$ doublet mode that we discovered. In this case we simultaneously observe two paired vortices with the same topological charge $q$, but with orthogonal polarizations at each point in the $k$-space. This can be seen from the polarization spirals associated with the two branches of the $E_2^{\mathrm{low}}$ mode in Figure 4e: the spirals for the lower (blue) and upper (green) branches are offset from each other by 90 degrees. The polarization vortex patterns measured for this SP-BIC are shown in Figure SM8 in the Supplementary material.

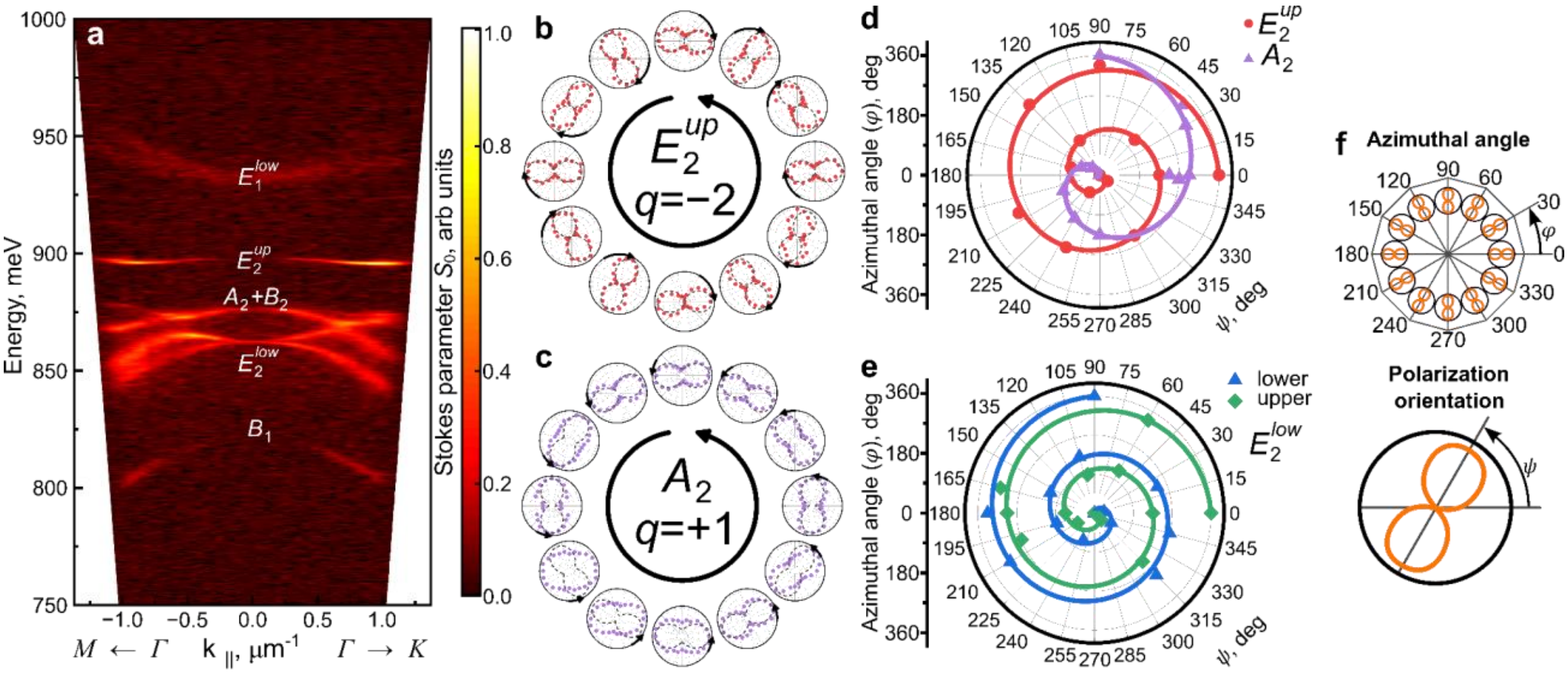

Figure 4 (a) Experimentally measured band diagram of PCS emissivity; (b) and (c) Polarization vortices measured for SP-BICs localized in the $E_2^{\mathrm{up}}$ and $A_2$ modes. The dots on the polarization figure-eights represent experimental data, the gray dashed lines show the results of numerical simulations;

(d) and (e) Polarization spirals demonstrating the change in radiation polarization (rotation of polarization figure-eights) when going around the $\Gamma$-point. Knowing the direction of rotation and the number of turns of the polarization spiral, one can unambiguously determine the topological charge of the vortex which forms around the BIC. Polarization spirals are shown for SP-BICs localized in the $E_2^{up}$, $A_2$ and $E_2^{low}$ modes; (f) Schematic diagram explaining the polarization spirals. Panel (a) shows direct measurements of the PCS emissivity band structure acquired without using polarization elements in the experimental setup.

As already noted, the symmetry and polarization of modes are directly related. Along the high-symmetry directions, the mode polarization is strictly defined, and can be only *s*- or *p*- polarization. Moreover, in the Brillouin zone, there are six equivalent $\Gamma \rightarrow K$ directions and six equivalent $\Gamma \rightarrow M$ directions. Consequently, when moving from one $\Gamma \rightarrow K$ or $\Gamma \rightarrow M$ direction to another, the polarization must be preserved. In this case, since the topological charge is directly related to the rotation of the polarization vector around the BIC, then, based on symmetry considerations and knowing the polarization of the mode in the directions of high symmetry, one can directly determine the possible topological charge of the SP-BIC.

As mentioned above, the modes of the PCS under study can be divided into two types: those with the same parity and polarization in the $\Gamma \rightarrow K$ and $\Gamma \rightarrow M$ directions (modes $A_1$, $A_2$, $E_1^{low}$ and $E_1^{up}$) and those with different symmetry (the different parity) and polarization (modes $B_1$, $B_2$, $E_2^{low}$, and $E_2^{up}$) (see also Figures 1e and 1f). It can be shown that for modes with SP-BICs at $\Gamma$-point, and with the same polarization in the $\Gamma \rightarrow K$ and $\Gamma \rightarrow M$ directions, the topological charge must be $q = +1$ (see Fig. 5a). Accordingly, for modes with SP-BICs observed at the $\Gamma$-point, which polarization is different in the $\Gamma \rightarrow K$ and $\Gamma \rightarrow M$ directions, the topological charge takes the value $q = -2$ (Fig. 5b). Thus, in PCSs with $\boldsymbol{C}_{6v}$ symmetry, SP-BICs localized at the $\Gamma$-point can only exhibit topological charge $q = +1$ and $q = -2$. BICs with $|q| > 2$ may exist at higher diffraction orders, but have not yet been observed experimentally. SP-BIC with topological charges $q = -1$ and $q = +2$ are forbidden in PCSs with $\boldsymbol{C}_{6v}$ symmetry (Figure 5c), as the corresponding polarization vortex would not be $\boldsymbol{C}_{6v}$ symmetric.

Notably, the $A_1$ and $A_2$ modes with SP-BIC at the $\Gamma$-point are not the only ones sharing the same polarization along high symmetry directions. The radiative modes $E_1^{low}$ and $E_1^{up}$ also exhibit the same symmetry along the $\Gamma \rightarrow K$ and $\Gamma \rightarrow M$ directions. This implies that for these radiative modes, polarization vortices with topological charge $q = +1$ should also appear around the $\Gamma$-point, yet without forming a BIC!

This is confirmed by numerical simulation shown in Figures 5d-e. Here we present calculations of the emissivity angular diagrams (Stokes parameter $S_0$), the degree of polarization, polarization ellipses and orientation of the polarization ellipses for three SP-BIC modes ($B_1$, $A_2$ and $E_2^{up}$) and radiative mode $E_1^{low}$, all calculations were performed for fixed energy values. For the radiative mode $E_1^{low}$, we present two diagrams: one calculated for the energy value of 954 meV, which coincides with the mode's eigenfrequency at the Γ-point, and the other calculated for the energy value of 973 meV, at which only one, upper branch of the mode is observed in the analyzed range of wave vectors.

The difference between topological charges $q = +1$ and $q = -2$ is clearly visible in the angular diagrams of polarization ellipses, especially when comparing the $A_2$ and $E_2^{up}$ modes. For the radiative mode $E_1^{low}$ at the $\Gamma$-point, we also observe a polarization vortex similar to the SP-BIC vortex of the $A_2$ mode. For all energies except the energy 954 meV (which is close to the eigenfrequency of the $E_1^{low}$ mode at the $\Gamma$-point), the positions of the maxima of the degree of polarization and the Stokes parameter $S_0$ coincide in $k$-space (Figure 5e). However, for the $E_1^{low}$ mode at the $\Gamma$-point (energy 954 meV) the opposite situation is observed. Since $E_1^{low}$ is a doublet mode, the mode degeneracy should be lifted away from the $\Gamma$-point, and two branches of the mode should appear: the upper one with $B$-symmetry and $s$-polarization, and the lower one with $A$-symmetry and $p$-polarization (see Figures 1e-d). At the $\Gamma$-point, the polarizations of both branches are orthogonal to each other, so their total degree of polarization must vanish.

There is a slight difference in the emissivities of the upper and lower branches of the $E_1^{low}$ mode at the $\Gamma$-point. At 954 meV, the lower ($p$-polarized) branch dominates to some extent. At 973 meV only the upper $s$-polarized branch is observed, and the orientation diagram of the polarization ellipses looks similar to the SP-BIC diagram observed for the $A_2$ mode. Thus, we observe the formation of a polarization vortex without BIC. In this case, the polarization at the vortex center (the $\Gamma$-point) is undefined, as is the case of BIC, meaning that there is a polarization singularity (V-point). However, unlike BIC, this singularity is radiative, indicating a new type of polarization vortex. To our knowledge, this type of vortex has not been previously discussed in the literature.

At an energy of 920 meV, where the lower branch of the $E_2^{up}$ mode is observed, we also see a small contribution of the upper branch to the ellipse orientation diagram near the Γ-point (visible as a small element in the center of the diagram), since at this energy the upper branch has a slightly higher emissivity compared to the lower one.

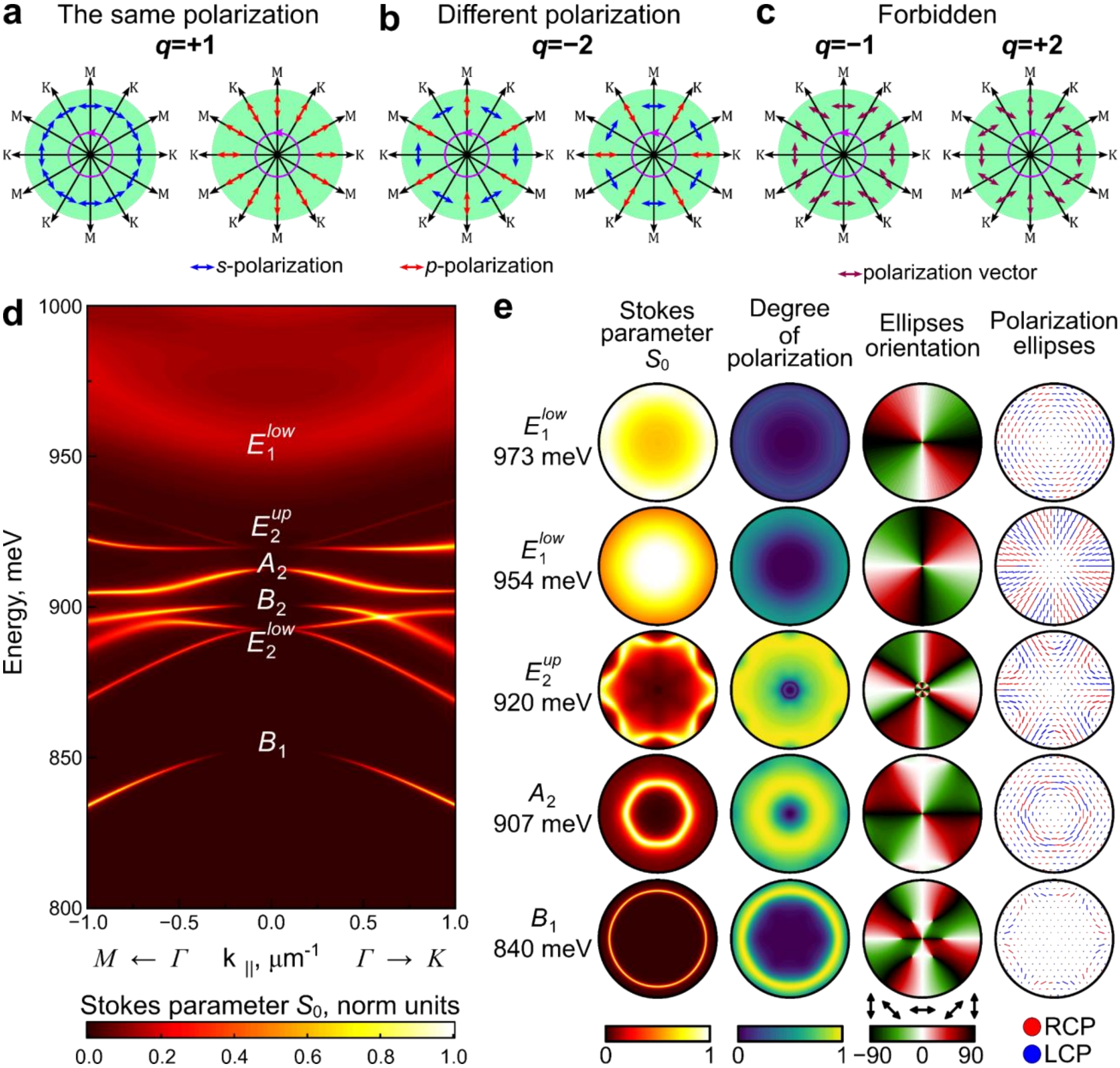

Figure 5 – Polarization vortices: (a) for modes with the same symmetry in the $\Gamma \rightarrow K$ and $\Gamma \rightarrow M$ directions; (b) for modes with different symmetry in the $\Gamma \rightarrow K$ and $\Gamma \rightarrow M$ directions; c) forbidden vortices. The arrows indicate the polarization vector in the high-symmetry direction. In panels (a) and (b), blue arrows correspond to *s*-polarization, and red arrows to *p*-polarization; d) Numerically simulated band structure of a PCS with the following lattice parameters: lattice period $a$=550 nm, hole diameter $d$=275 nm, etching depth $h_{etch}$=250 nm and Si waveguide thickness $h_{Si}$=335 nm; (e) Angular diagrams of the Stokes parameter $S_0$, degree of polarization, rotation angle of polarization ellipses, and polarization ellipses themselves calculated within the range of wave vectors up to $|k_{\|}|$=1 μm$^{-1}$ (the radius of the circles in which the diagrams are plotted) for PCS in the (d). The simulations were performed at fixed energies: 840, 907, 920, 954 and 973 meV. The ellipse sizes are proportional to the intensity of the polarized emissivity. Higher-resolution polarization ellipse diagrams are shown in Figure SM9 of the Supplementary materials.

As discussed above, in PCSs with $\boldsymbol{C}_{6v}$ lattice symmetry, only SP-BICs with topological charges $q$=+1 and $q$=−2 can exist at the $\Gamma$-point. BICs with other topological charges are forbidden by symmetry. Here it is interesting to consider BIC of a different nature, localized outside the $\Gamma$-point. As our recent studies have shown, outside the $\Gamma$-point in PCS with $\boldsymbol{C}_{6v}$ lattice symmetry, the Friedrich-Wintgen BICs (FW-BICs) can also be observed.[75]

It can be shown that FW-BIC are not subject to the same symmetry restrictions as SP-BIC. Consider, for example, the FW-BIC arising on the upper branch of the $E_1^{low}$ mode in the $\Gamma \rightarrow M$ direction (Figure 6a). This FW-BIC results from the avoided crossing of two modes of the same $B$ symmetry, namely the upper branch of the $E_1^{low}$ mode and the lower branch of the $E_2^{up}$ mode.[75] This is evidenced by the same sign of the Stokes parameter $S_1$ of these two modes (see Figure 6b). The calculated Stokes parameters $S_0$ and the degree of mode polarization near the FW-BIC are shown in Figures 6c and 6d. The yellow stars in these figures indicate the position of FW-BIC in $k$-space. At the FW-BIC point, both the Stokes parameter $S_0$ and the degree of polarization are minimal, which is similar to the behaviour of the SP-BICs. Figures 6e and 6f show the map of polarization ellipses and their orientation diagram near the FW-BIC. From these figures, the topological charge of the observed FW-BIC is $q = -1$.

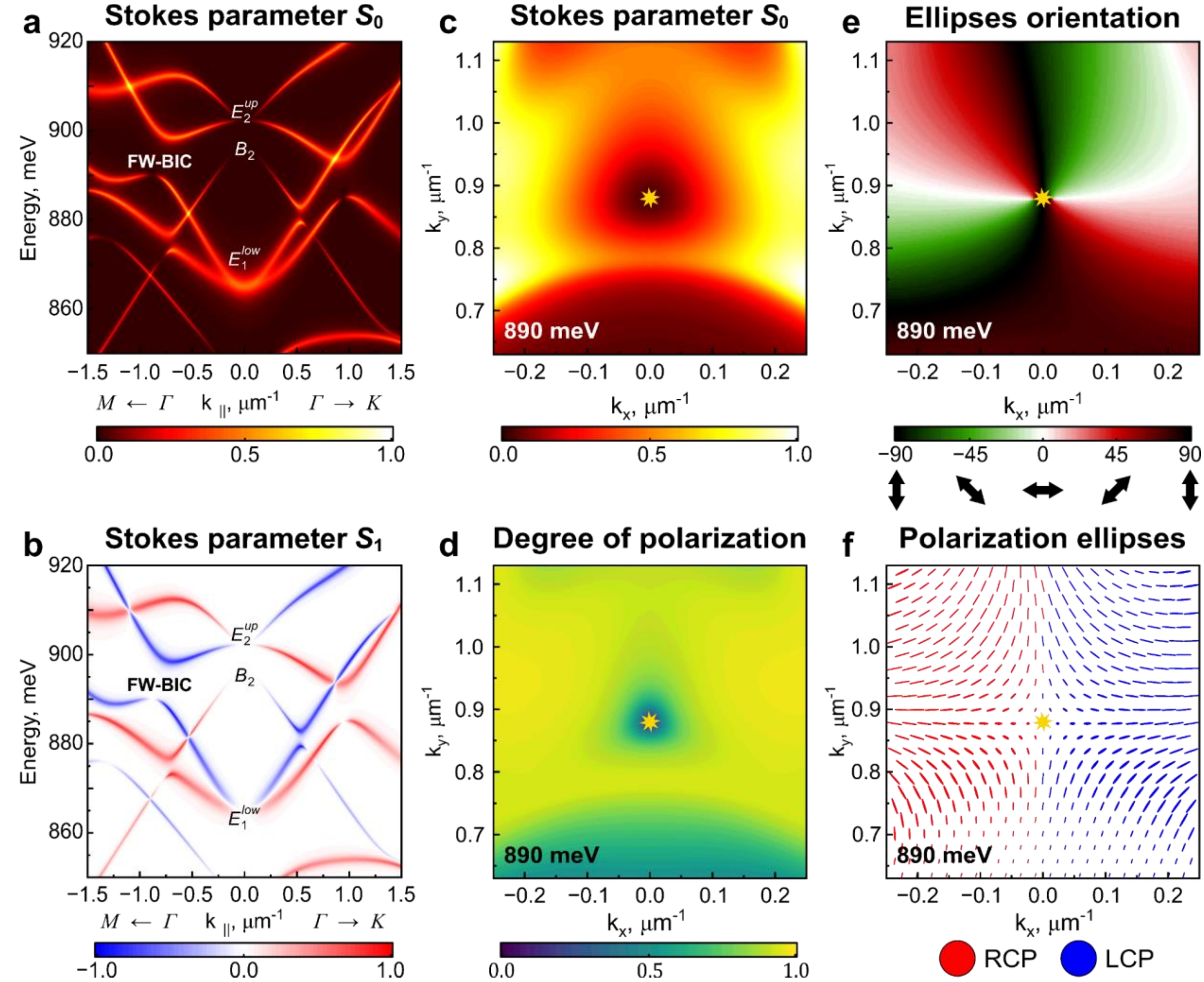

Figure 6 (a) and (b) Calculated band diagram of PCS, presented in Stokes parameters S0 (a) and S1 (b). (c,d,e,f) Plotted in the wave vector plane: PCS emissivity (Stokes parameter S0) (c), degree of polarization (d), orientation diagram of polarization ellipses (e), map of polarization ellipses (f). All data were obtained for a fixed energy of 890 meV. The diagrams in panels (c)-(f) were calculated near the FW-BIC marked with a yellow star. The ellipse sizes are proportional to the intensity of the polarized emissivity. In this figure, PCS with the following lattice parameters was analyzed: lattice period $a$ = 550 nm, hole diameter $d$ = 220 nm, hole etching depth $h_{etch}$ = 135 nm, the thickness of silicon waveguide layer $h_{Si}$ = 335 nm.

**Materials and methods**

Below, we provide data for the sample that was experimentally studied in this work. The parameters of the samples for which theoretical calculations were performed are listed in the text of the article and in the captions of the figures.

*Sample fabrication*: The PCS studied in this work was fabricated on a structure with self-assembled Ge(Si) nanoislands grown by molecular beam epitaxy on an SOI wafer with a 70-nm-thick Si device layer and a 2-μm-thick buried oxide (BOX) layer. The structure contained a 50-nm-thick silicon buffer layer, five layers of Ge(Si) nanoislands separated by 17-nm-thick Si layers, and a 135-nm-thick Si cap layer. The total thickness of the structure above the buried oxide layer was 335 nm. The Ge(Si) nanoislands were grown at a temperature of 600 °C by the depositing of 10 monolayers of Ge.

The PCS was fabricated using electron beam lithography and plasma-chemical etching. At the first step, a PCS pattern was formed in a PMMA resist by electron-beam lithography. This pattern served as an etching mask. Anisotropic etching of the PCS was performed using ICP plasma-chemical etching in the $SF_6/C_4F_8$ mixture of gases. In this work, we studied a PCS with a hexagonal lattice of air holes in the Si layer. The fabricated PCS has the following geometric parameters: lattice period $a$=575 nm, hole diameter is $d$=277 nm, and etching depth $h_{etch}$=250 nm. The size of the PCS was 50×50 μm and contained more than 86×86 photonic crystal lattice periods.

*PL measurements setup*: We employed two experimental techniques to study the light-emitting properties of the PCS: a standard micro-PL setup, and an original angle-resolved PL setup.[71,73,75,79]

The first technique enables measurements with high spatial resolution. In this setup, the PL signal is excited and collected using the same objective. We used microscope objective with ×10 magnification (Mitutoyo M Plan Apo NIR 10x, NA=0.26) and collection angle $2\vartheta = 30°$.

The second technique allows for measurements with angular resolution and experimental study of the PCS band structure. Angular resolution in the measuring schema was provided by introducing a diaphragm into the parallel beam formed by the objective. Figure S10 in Supplementary materials shows this measurement setup, which we also discussed in detail in our earlier work.[75] To analyze the polarization characteristics of the radiation, we modified this scheme by adding two polarization elements to the optical path: an achromatic quarter-wave plate (QWP) and a polarization analyzer (linear polarizer). Sequential installation of the analyzer and the QWP in the optical path makes it possible to measure the Stokes parameters $S_1$ and $S_2$. The sequential installation of the QWP and the analyzer with an angle of 45 degrees between their optical axes enables it to measure the Stokes parameter $S_3$. This sequence of installation of the QWP and analyzer is necessary to take into account the absorption of QWP when calculating the experimentally measured degree of polarization. In addition, in this setup we use so-called "side-pumping" to excite the PL signal. This approach allows us to excite the entire photonic crystal slab and increase the intensity of the recorded PL signal by an order of magnitude without increasing the power density of the exciting radiation.[79]

In both setups, the PL signal was excited by a continuous-wave Nd:YAG laser with a wavelength of 532 nm. A Bruker IFS 125HR high-resolution Fourier spectrometer with a cooled germanium detector was used to record the spectra.

*Theoretical method*:

To theoretically study the optical behavior of the PCS with Ge nanoislands, we use a Fourier modal method (FMM) in the scattering matrix form,[80] also known as rigorous coupled-wave analysis (RCWA).[81] In the Fourier decomposition of electromagnetic fields, to preserve the $\boldsymbol{C}_{6v}$ symmetry of the structure, we choose the corresponding set of Fourier harmonics in the reciprocal space. The total number of harmonics is chosen to be $N_g = 121$ that ensures the convergence of our numerical scheme. As a result, we construct the $4N_g \times 4N_g$ dimensional scattering matrix $S(\omega,k)$ which contains full optical information of our photonic crystal slab. Here $\omega$ and $k$ denote the frequency of electromagnetic oscillations and in-plane wave vector respectively.

The emissivity of Ge(Si) nanoislands is modeled using Kirchhoff's law, whereby we calculate the absorptivity of an incident plane wave at a given frequency, direction, and polarization state instead of directly computing the emissivity. Since the PCS material (silicon) is transparent in the frequency range of interest, we introduce a small imaginary part ($\varepsilon'' = 0.01i$) to its dielectric permittivity, applied uniformly across the entire emitting layer. The emissivity is then proportional to the computed absorptivity, with a proportionality coefficient that is independent of the PCS geometry and incident polarization.

To estimate the Stokes parameters, we compute the emissivity for different polarization states and then apply the following standard expressions:

$$S_0 = I_x + I_y,$$
$$S_1 = I_x - I_y,$$
$$S_2 = I_a - I_b,$$
$$S_3 = I_r - I_l.$$

where the subscripts $x$, $y$, $a$, $b$, $r$, and $l$ denote polarizations corresponding to the following Jones vectors in the $s$–$p$ basis:

$$J_x = \begin{bmatrix} 1 \\ 0 \end{bmatrix},$$
$$J_y = \begin{bmatrix} 0 \\ 1 \end{bmatrix},$$
$$J_a = \frac{1}{\sqrt{2}}\begin{bmatrix} 1 \\ 1 \end{bmatrix},$$
$$J_a = \frac{1}{\sqrt{2}}\begin{bmatrix} 1 \\ -1 \end{bmatrix},$$
$$J_r = \frac{1}{\sqrt{2}}\begin{bmatrix} 1 \\ -i \end{bmatrix},$$
$$J_l = \frac{1}{\sqrt{2}}\begin{bmatrix} 1 \\ i \end{bmatrix}.$$

The Stokes parameter $S_0$ corresponds to polarization-average emissivity. In figures, the $S_0$ maps are normalized to their maximum value, while the $S_1$, $S_2$ and $S_3$ maps are normalized to the unnormalized $S_0$ parameter.

**Conclusion**

Thus, in this work we have demonstrated the possibility to observe in photonic-crystal slabs with a hexagonal lattice of holes a wide spectrum of polarization states of radiation, extending from linear to circular polarizations and polarization vortices with different topological charges. The observed polarization features are associated with the specificity of the modes, the features of their propagation and the phenomena of intermode interaction that occur in such photonic-crystal slabs. We have shown that along the high-symmetry $\Gamma \to M$ and $\Gamma \to K$ directions in the first Brillouin zone, the modes of a PCS are linearly polarized. For all modes in these directions, only the Stokes parameter $S_1$ is nonzero, which allows us to separate PCS modes by polarization (*s* or *p*) and symmetry. Beyond the $\Gamma$-point of the Brillouin zone, modes of *A*-type symmetry can emit only *p*-polarized light, and modes of *B*-type symmetry can emit only *s*-polarized light, allowing for a unique classification of the modes observed in the experiment. The symmetry of modes also determines the topological charge of the polarization vortices formed in such photonic-crystal slabs. Outside the high-symmetry directions of the first Brillouin zone, all three Stokes parameters characterizing PCS emission can be non-zero; moreover, one can also observe the C-points where the emission of PCS is circular polarized. In addition to linear and circular polarizations, we also demonstrated the possibility to observe polarization vortices in PCSs with a hexagonal lattice of holes. We simultaneously observed 4 vortices with topological charges $q$=+1 and $q$=−2, localized on SP-BICs. It is shown that SP-BICs formed in photonic crystal slabs with $\boldsymbol{C}_{6v}$ lattice symmetry have vortices only with such topological charges, whereas for FW-BICs a polarization vortex with a topological charge $q$=−1 can be observed. We also discovered new types of polarization vortices. These are polarization vortices characterized by two mutually orthogonal polarizations with the same topological charge, and polarization vortices formed without the presence of BIC. The former have been observed for SP-BICs formed on doublet modes, while the latter are characteristic of radiative modes. A vortex without a BIC is formed around the polarization singularity of the radiative mode at the $\Gamma$-point of the Brillouin zone. The polarization phenomena we have discovered open up new possibilities for creating compact radiative sources with selected polarization, which can be used in integrated and quantum photonics, information processing and transmission circuits, sensing applications, and more.

## Acknowledgements

Experimental investigation of luminescent properties and polarization features of photonic crystals was funded by the Russian Science Foundation (grant # 25-12-00367). The growth of Si/SiGe structures and photonic crystal formation have been performed under state assignment of IPM RAS (FFUF-2024-0019). Theoretical part of the work has been funded by Russian Science Foundation (grant # 25-12-00454). A.A.B. and R.K.N. acknowledges the National Natural Science Foundation of China (Grant No. W2532010) and the Priority 2030 Academic Leadership Program. Authors acknowledge D.A. Chermoshentsev for providing an achromatic quarter-wave plate for measuring Stokes parameters.

## Author Contribution

A.V.P. contributed to the development of the measurement methodology and performed the research presented in this paper, including processing the obtained data and numerical modeling of the experiment. A.V.P. prepared the draft and final versions of the manuscript. M.V.S. contributed to the development of the measurement methodology, supervised the experimental studies, and participated in writing and editing the manuscript. N.A.G., A.A.B., S.G.T. and R.K.N. took an active part in the discussion and interpretation of the obtained results, their numerical modeling, as well as in writing and editing the manuscript, and preparing the manuscript for publication. M.V.Sh. contributed to the sample growth and fabrication. A.V.N. supervised the growth and fabrication of samples and took an active part in the discussion of the results. S.A.D. supervised the overall project, developed and implemented the numerical modeling methods used in this work, took an active part in discussing the results, writing and editing the manuscript, and preparing the manuscript for publication.

**Data availability**

The data that support the findings of this study are available from the corresponding author upon reasonable request.

**Conflict of interest**

The authors declare no competing interests.